\documentclass[preprint,12pt]{elsarticle}

\usepackage{amssymb}
\usepackage{amsmath}
\def\empile#1\over#2{\mathrel{\mathop{\kern 0pt#1}\limits_{#2}}}
\def\p{{\boldsymbol p}}

\def\r{{\boldsymbol r}}

\begin{document}

\begin{frontmatter}



\title{{\bf Isotropization in heavy ion collisions\\ in the 2PI formalism\footnote{Talk given at XQCD 2025, Wroclaw, Jul 2-4, 2025.}}}


\author{Fran\c cois Gelis} %% Author name

\affiliation{organization={Institut de Physique Théorique},%Department and Organization
            addressline={Université Paris-Saclay, CEA/DRF},
            city={Gif sur Yvette},
            postcode={91191},
            state={},
            country={France}}

%% Abstract
\begin{abstract}
An important question in heavy-ion collisions is how the initial
far-from-equilibrium medium evolves and thermalizes while it undergoes
a rapid longitudinal expansion. In this presentation, we show how
to use the two-particle irreducible (2PI) effective action to address
this question, focusing on $\phi^4$ scalar theory truncated at three
loops. We present numerical results for quantities such as the
occupation number, the thermal mass, and the number density.
\end{abstract}

%% Keywords

\end{frontmatter}



\section{Introduction}
\begin{figure}[htbp]
  \centering
  \includegraphics[width=100mm]{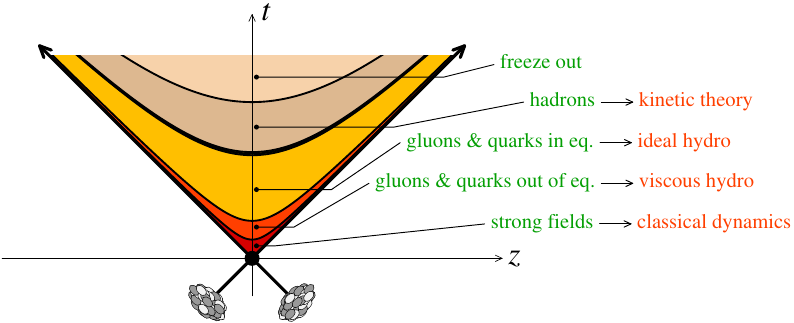}
  \caption{\label{fig:HIC}Main stages of a heavy ion collision.}
\end{figure}

Heavy ion collisions at relativistic energies are used to probe
experimentally the deconfined state of nuclear matter. These
collisions, although in principle described by quantum chromodynamics (QCD),
are usually described using several effective descriptions that hide a
lot of the microscopic details, as highlighted in Figure
\ref{fig:HIC}.
Particularly important among these tools is relativistic
hydrodynamics. Its effectiveness in describing the expansion of the
matter produced in heavy ion collisions
\cite{Gale:2013da} stems from two facts. Firstly
the viscosity of the quark gluon plasma is extremely small,
$\eta/s\sim 0.1$, and secondly, the pressure tensor is not too
anisotropic. The latter point has proven difficult to justify from the
underlying microscopic description.

A description of the initial stages of a heavy ion collision, directly
inspired from QCD, is provided by the color glass condensate
(CGC) \cite{Iancu:2000hn,Gelis:2010nm}. This is an effective theory in which one separates the partonic
degrees of freedom inside the colliding nuclei into hard and soft,
according to their longitudinal momentum. The hard partons are
described globally as a density of color charges, that produces a
color current pointing along the collision axis. Because they are
fast, these modes appears highly boosted in the lab frame, and
therefore have no evolution thanks to time dilation. In contrast, the
soft degrees of freedom as described using the usual gauge fields.

\begin{figure}[htbp]
  \centering
  \includegraphics[width=100mm]{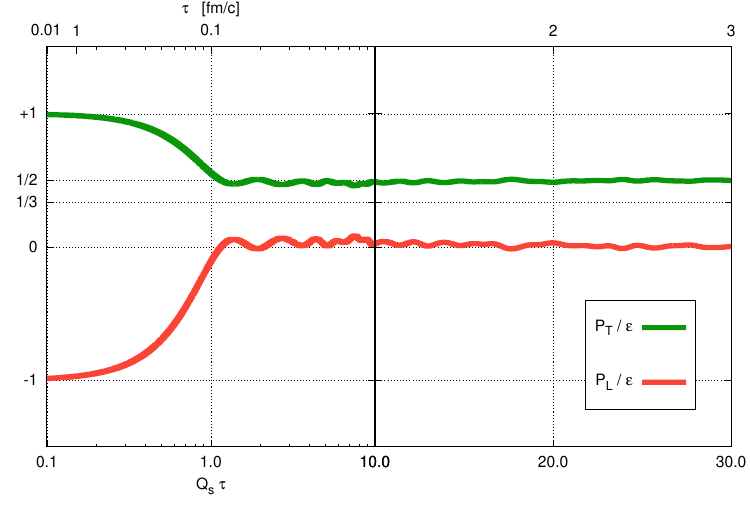}
  \caption{\label{fig:ratio}Leading order transverse and longitudinal pressures versus time.}
\end{figure}

The leading order in the CGC amounts to solving the classical
Yang-Mills equations with the color currents of the projectiles as
sources, and then feeding the solution into observables. In this way,
one can evaluate the components of the energy-momentum tensor. The
transverse and longitudinal pressures, normalized by the energy
density, are shown as a function of proper time in Figure \ref{fig:ratio}.
At very early times after the collision, the longitudinal pressure is
negative \cite{Lappi:2006fp,Fukushima:2011nq} and the system cannot be described as a collection of
quasiparticles. At times larger than the inverse of the saturation
momentum, the longitudinal pressure is no longer negative, but remains
much smaller than the transverse one. The ratio $P_{_L}/P_{_T}$ goes
to zero, and the system becomes more and more anisotropic, consistent
with a free streaming system.

Isotropization, if it happens, comes from effect beyond the leading
order. In the CGC, the next-to-leading order shows some trend towards
isotropization, but is plagued with secular divergences due to
instabilities in the solutions of linearized Yang-Mills equations \cite{Romatschke:2005pm}. An
all-orders resummation is provided by kinetic theory, but the kinetic
approximation requires a number of assumptions,
\begin{itemize}
\item long-lived quasiparticles,
\item moderate occupation number,
\item moderate gradients,
\end{itemize}
that are not realized in the underlying theory at times around the
inverse saturation scale.

\section{2PI effective action}
Kinetic theory can be derived from the underlying quantum field
theory, by first writing the Dyson-Schwinger equation that resums a
self-energy $\Sigma$ on the propagator,
\begin{align}
  G=G_0+G_0\Sigma G,
  \label{eq:DS}
\end{align}
which is an exact equation as long as we do not truncate the
self-energy $\Sigma$. Kinetic theory is obtained from
eq.~(\ref{eq:DS}) by truncating $\Sigma$ (usually at the lowest loop
order that gives scatterings), by doing an expansion in gradients, and
by doing a quasi-particle approximation (this is a property about the
behavior of some propagators, that are assumed to be proportional to
the spectral function).

The two-particle irreducible (2PI) approximation \cite{Berges:2004yj}
consists in solving directly eq.~(\ref{eq:DS}) with just a truncation
of $\Sigma$, but without the other two approximations that lead to a
kinetic equation. A more formal derivation of the equations of motion
in the 2PI formalism can be obtained from the 2PI effective
action. Let us recall that the ordinary (one-particle irreducible)
effective action is a generalization of the classical action that
includes all quantum corrections, under the assumption that we know
the expectation value of the field, $\varphi$ (knowing the expectation
value of the field allows to hide in $\varphi$ infinite sets of
Feynman graphs, and thus simplifies the perturbative expansion). The
2PI effective action takes this logic a step further, by assuming that
we know both the expectation value of the field, $\varphi$, and the
exact propagator $G$. The remaining perturbative expansion simplifies
even more, because additional infinite sets of Feynman graphs can be
hidden in $G$.

For the $\phi^4$ scalar field theory that we consider in the rest of
this presentation \cite{Gelis:2024iar}, the 2PI effective action reads
\begin{align}
  \Gamma[\varphi,G]=S[\varphi]-\frac{i}{2}{\rm tr}\,\big(\log G\big)
  +\frac{i}{2}{\rm tr}\,\big(G_0^{-1}\,G\big)+\Phi[\varphi,G].
\end{align}
In this expression, $G_0$ is the free propagator in the background
field $\varphi$, that obeys
\begin{align}
  G_0^{-1} =i\left(\square_x+m^2+\tfrac{g^2}{2}\varphi^2(x)\right)\delta(x-y),
\end{align}
$G$ is the exact propagator, $S[\varphi]$ is the classical action and
$\Phi[\varphi,G]$ is the sum of all {{2PI} vacuum graphs}. The
equations of motions that drive the evolution of $\varphi$ and $G$ are
\begin{align}
  \frac{\delta \Gamma}{\delta \varphi(x)}=0,\quad
  \frac{\delta \Gamma}{\delta G(x,y)}=0,
\end{align}
or more explicitly,
\begin{align}
  &(\square_x\!+\!m^2)G(x,y)+\tfrac{g^2}{2}\varphi^2(x)G(x,y)=-i\delta(x-y)
  +\int d^4z\,{\frac{\delta \Phi}{\delta G(x,z)}}\,G(z,y),\nonumber\\
  &(\square_x\!+\!m^2) \varphi(x) + \tfrac{g^2}{6}\varphi^3(x) +\tfrac{g^2}{2}G(x,x)\varphi(x)=\frac{\delta\Phi}{\delta\varphi(x)}.
  \label{eq:eom}
\end{align}
These equations are still exact, but impractical since the functional
$\Phi$ is an infinite sum of 2PI graphs. In this study, we have
truncated this functional at order $g^4$:
\setbox1\hbox to 40mm{\includegraphics[width=40mm]{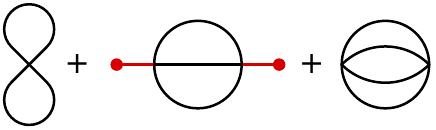}}
\begin{equation}
  i\Phi[\varphi,G]\empile{=}\over{\mbox{\scriptsize order $g^4$}}\;\;
  \raise -5mm\box1.
  \label{eq:PHI}
\end{equation}
Note that the equations of motion (\ref{eq:eom}) are expressed in the
Schwinger-Keldysh formalism, i.e., all time integrations are extended
to the closed time path. In practice, one needs to decompose the
propagator in two components\footnote{This decomposition does not
change the number of degrees of freedom in the equations of motion,
but replaces the complex valued propagator $G$ by two real valued
propagators $\rho$ and $F$.}, the spectral function $\rho$, and the
statistical propagator $F$ (that contains information about the
particle content of the system).

\section{Renormalization}
Like the ordinary perturbation, the 2PI formalism contains ultraviolet
divergences. However, the renormalization of the 2PI equations of
motion requires non-perturbative counterterms in order to cope with
the fact that infinitely many ordinary Feynman graphs are being resummed.

The most prominent divergence comes from a tadpole, of order $g^2$,
that diverges quadratically in the ultraviolet cutoff. Since the
tadpole is momentum independent, it requires only a mass counterterm
and no field renormalization.

In addition, some of the resummed graphs contain 1-loop corrections to
the vertex, starting at order $g^4$, that depend logarithmically on
the ultraviolet cutoff.

It turns out that, because of the resummations performed by the 2PI
equations of motion, one cannot renormalize the mass independently of
the vertex. In this work, we have performed a renormalization that
would be exact in the Hartree approximation, i.e., if we had kept only
the first term in eq.~(\ref{eq:PHI}). At this level, the bare and
renormalized mass and coupling constant are related by
\begin{align}
  &m_{_B}^2=m_{_R}^2+\delta m^2 ,\quad
  \delta m^2 = -\frac{g_{_R}^2+\delta g^2}{2}\int_k F_{\rm vac}(k,m_{_R}^2),
\end{align}
\begin{align}
  g_{_B}^2\equiv  g_{_R}^2+\delta g^2
    =
    \frac{g_{_R}^2}{1-\tfrac{g_{_R}^2}{2}\int_k  F_{\rm vac}(k,m_{_R}^2) \rho_{\rm vac}(k,m_{_R}^2)}.
\end{align}
This prescription consistently subtracts all divergences in the
Hartree truncation, but leaves unsubtracted some divergences due to
the 2-loop self-energies. With a coupling $g^4=500$, the $\delta m^2$
combing from the tadpole is the largest subtraction, $\delta g^2$
provides a $10-15$\% correction, while the $\delta m^2$ coming from the
2-loop self-energies is much smaller, as well as the field
renormalization, both at the percent level.

\section{Numerical implementation}
Even more than the raw computational power, the memory needs are a
major concern for solving the 2PI equations of motion, because of the
need to store the past propagators (in principle over the whole
history of the system). For instance, to store the past propagators
over $N_t=1000$ timesteps, for a spatial volume $N^3=500^3$, in
simple precision, one would need of the order of $10^9$ TeraBytes.

In order to bring the memory footprint down to a more realistic
figure, we assume that the system under consideration posseses some
symmetries:
\begin{itemize}
\item translation and rotation invariance in the plane transverse to the collision axis,
\item invariance under boosts along the collision axis (in order to
  make its manifestation simpler, we use proper time and rapidity as
  coordinates, instead of $x^0,x^3$),
\item collision of identical projectiles, so that the propagators have
  a definite parity (even or odd).
\end{itemize}
With these symmetries, the storage of the propagators scales as
$\tfrac{1}{2}N_t^2N_\perp N_\eta$, where $N_\perp$ is the number of
grid points in the radial direction of the transverse plane, and
$N_\eta$ is the number of grid points in rapidity. With $N_t=1000,
N_\perp=N_\eta=500$, one now needs only $0.5$ TeraBytes, which is
perfectly achievable.

Note that, even though the computations themselves are not the major
bottleneck, the problem at hand is easily parallelizable and there is
definite gain in using GPUs instead of traditional CPUs. The advantage
comes both from their performance in massively parallel computations,
and in their high memory bandwidth (provided most of the necessary
data is resident on the memory of the GPUs).

\section{Results}
The free parameters one must choose in order to initialize the
evolution of the system are the following:
\begin{itemize}
  \item initial time $\tau_0$,
  \item initial occupation number $f(\tau_0,p_\perp,\nu)$,
  \item initial field $\varphi(\tau_0)$ and derivative $\dot{\varphi}(\tau_0)$,
  \item renormalized mass $m_{_R}$,
  \item renormalized coupling constant $g^2_{_R}$.
\end{itemize}
In the plots presented below, we have used $g_{_R}^2\approx 22$, which
is a moderate coupling\footnote{At this coupling, the shear viscosity
to entropy ratio is $\eta/s\sim 25 \gg 1$.} for the $\phi^4$ theory,
and the initial field expectation value was set to zero $\varphi=0$
(subsequently, it stays null at all times). By solving the 2PI
equations of motion, one obtains $\varphi, F, \rho$. In order to
extract some quantities from the propagators, we assumed a
quasi-particle ansatz in which the propagators keep the same form as in
the vacuum, but with a different particle distribution and a different
mass (due to in-medium effects). For instance, within this ansatz, the
occupation number $f$ is obtained as
\begin{align}
    \tfrac{1}{2}+f(\p; \tau)
    \empile{\approx}\over{\tau=\tau'}
      \tau\sqrt{F(\partial_\tau\partial_{\tau'}F)-(\partial_\tau F)(\partial_{\tau'}F)}.
\end{align}
(Note that, in order to obtain the momentum dependence of $f$, we must
Fourier transform the propagators on their spatial coordinates
$\r_\perp$ and $\eta$.) Likewise, one may extract an effective mass
from the propagators by studying their oscillation period in $\tau$.

\begin{figure}[htbp]
  \centering
  \includegraphics[width=100mm]{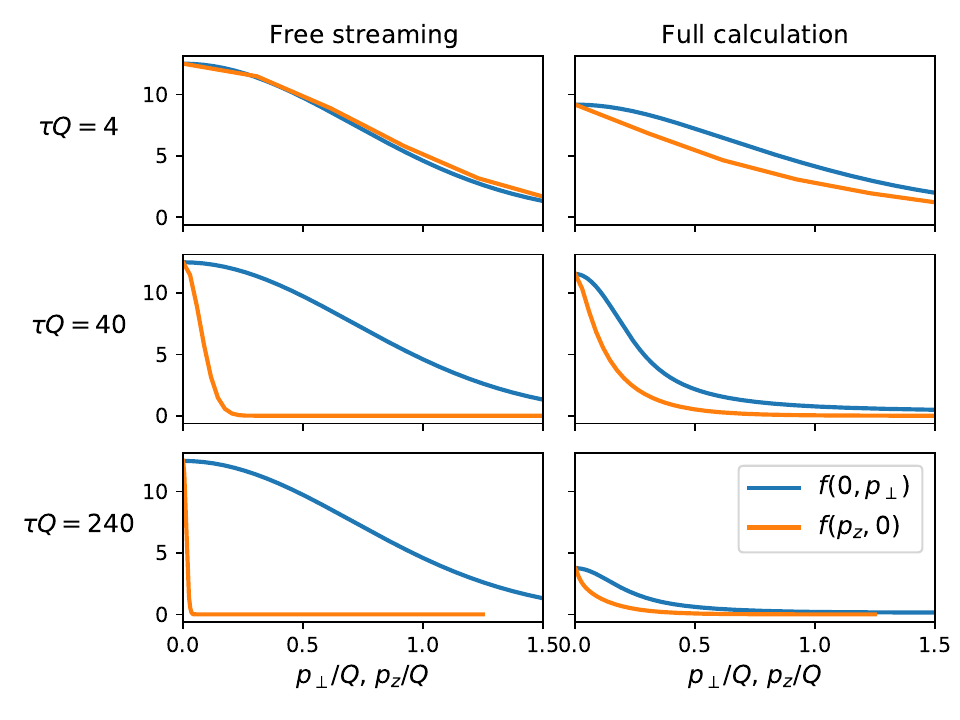}
  \caption{\label{fig:density}Occupation number as a function of
    $p_\perp$ and $p_z$, at three different times in the
    evolution. Blue: $p_\perp$ dependence. Orange: $p_z$ dependence. Left: Hartree approximation. Right: with 2-loop
    self-energy.}
\end{figure}

In figure \ref{fig:density}, we show the extracted particle density. 
The plots in the left column are for a computation in the Hartree
approximation, while the right column had also the 2-loop
self-energy. The blue curves show the dependence on $p_\perp$, and the
orange curves show the $p_z$ dependence. In the Hartree truncation, we
observe that the support of the $p_z$ distribution shrinks very
rapidly and the system becomes more and more anisotropic, consistent
with free streaming (the system has longitudinal expansion, but no
scatterings). In contrast, when the 2-loop self-energy is included,
the anisotropy in the system remains moderate, a sign that the 2-loop
graphs bring scattering processes that are able to fight against the
expansion.

\begin{figure}[htbp]
  \centering
  \includegraphics[width=100mm]{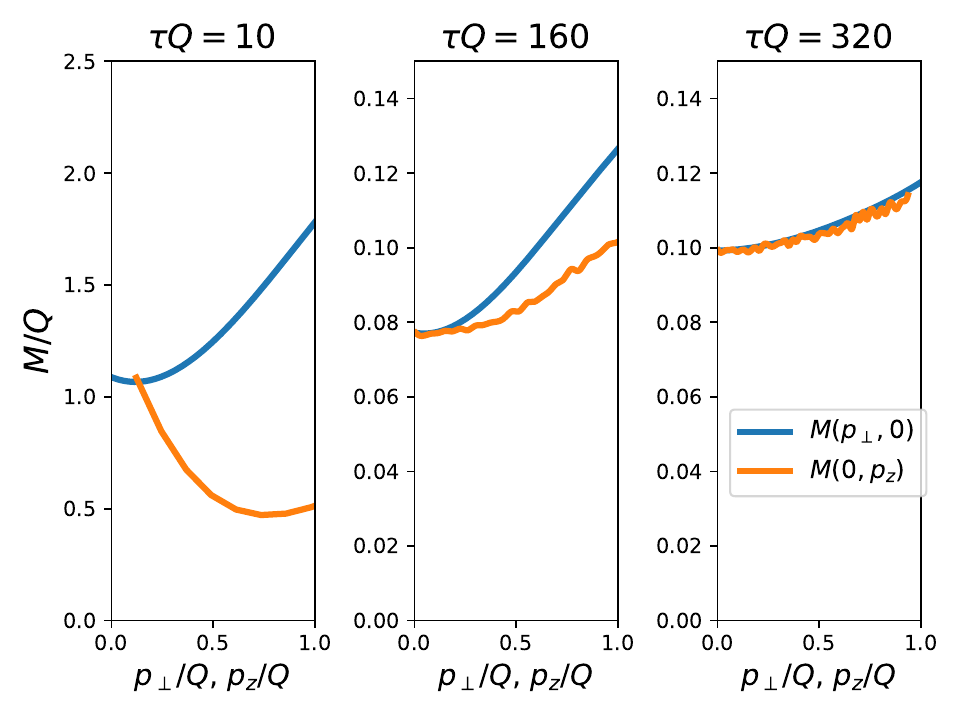}
  \caption{\label{fig:mass}Effective mass extracted from the propagator, at three proper times. Blue: $p_\perp$ dependence. Orange: $p_z$ dependence.}
\end{figure}

In figure \ref{fig:mass}, we show the effective mass extracted from
the propagators, at three times during the evolution of the
system. More than just a mass, the extracted object is a self-energy
that depends on $p_\perp$ and $p_z$.
At the beginning of the evolution, the momentum dependences in
$p_\perp$ and $p_z$ are very different, reflecting the fact that the
internal degrees of freedom of the system have a very anisotropic
distribution. Then, the $p_\perp$ and $p_z$ dependences of the
self-energy become more and more similar, to be almost identical at
the latest time considered. We interpret this as another sign of
isotropization.

\begin{figure}[htbp]
  \centering
  \includegraphics[width=100mm]{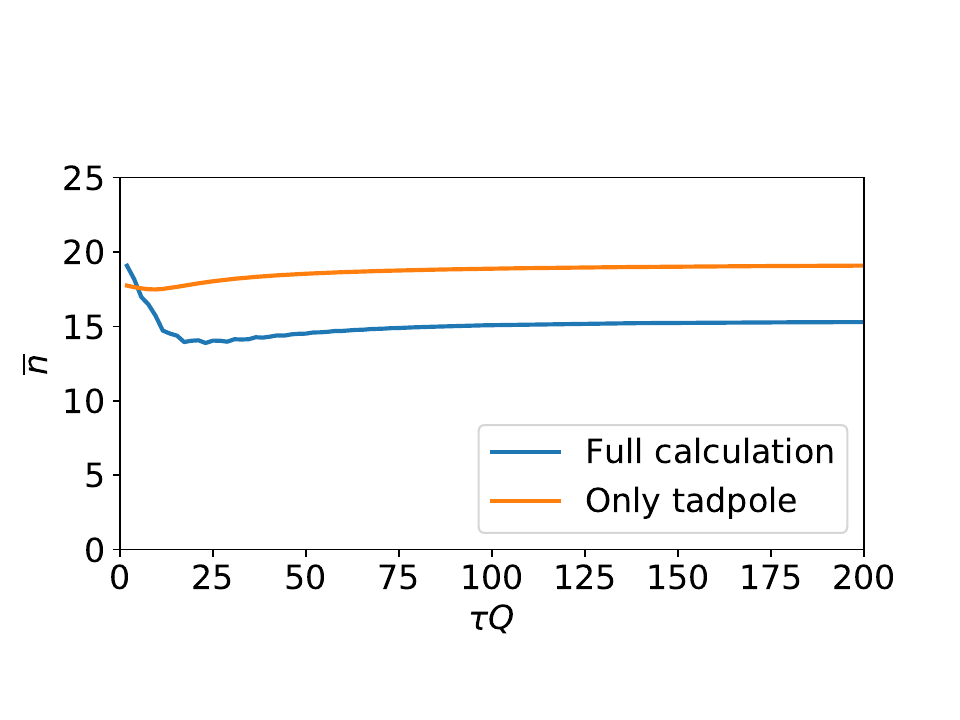}
  \caption{\label{fig:number}Evolution of $\tau n(\tau)$, in the
    Hartree approximation (orange curve) and with the 2-loop
    self-energy (blue curve).}
\end{figure}

In figure \ref{fig:number}, we have integrated the extracted
occupation number of all momenta, in order to construct the number of
particles per unit volume, $n(\tau)$. The plots in this figure show
$\overline{n}(\tau)\equiv \tau n(\tau)$, as a function of proper time,
in the Hartree approximation (orange curve) and when the 2-loop
self-energy is also included (blue curve).
When only the tadpole is taken into account, this quantity is almost
constant in time, in agreement with a system whose only scatterings
are elastic. In contrast, when we include the 2-loop self-energy, the
number density displays a clear drop at the beginning of the
evolution. We interpret this as the possibility to have $3\to 1$
processes with off-shell particles, allowed in the 2PI framework when
we include the 2-loop self-energy. This is at odds with kinetic
theory, where the same self-energy only allows number preserving $2\to
2$ processes.  In the present case, these off-shell processes tend to
decrease the number density because we started from an overpopulated
system, that had more particles than an equilibrated system with the
same energy density.

\section{Summary}
In this work, we have applied the 2PI formalism truncated at order
$g^4$ to the description of a longitudinally expanding system of
massive scalar fields (with an initial condition such that the field
expectation value is null at all times). From the time dependence of
the propagators, we have extracted the particle occupation number,
number density, and a momentum dependent effective mass. These
quantities show that the system approaches an almost isotropic state,
despite the longitudinal expansion. Moreover, the effect of number
changing off-shell processes -- that tend to tame an initial
overpopulation in the system -- is also visible.

\paragraph{Acknowledgements}
This work was granted access to the HPC resources of IDRIS under the
allocation 2023-AD010514330 made by GENCI. This work is supported in
part by the GLUODYNAMICS project funded by the ``P2IO LabEx (ANR-
10-LABX-0038)'' in the framework ``Investissements d’Avenir''
(ANR-11-IDEX-0003-01) managed by the Agence Nationale de la Recherche
(ANR), France.

%\bibliography{refs} \bibliographystyle{unsrt}

\end{document}